\documentclass[prb,onecolumn,12pt]{revtex4-1}
\usepackage{epsfig}
\usepackage{amssymb}
\usepackage{threeparttable}
\usepackage{graphicx}
\usepackage{natbib}
\usepackage{longtable}
\usepackage{txfonts}
\usepackage{color}

\begin{document}
\title{Atomically sharp 1D SbSeI, SbSI and SbSBr with high stability and novel properties for microelectronic, optoelectronic, and thermoelectric applications}
\author{Bo Peng$^1$, Ke Xu$^1$, Hao Zhang$^{1,4*}$, Zeyu Ning$^1$,  Hezhu Shao$^{2}$, Gang Ni$^1$, Hongliang Lu$^3$, Xiangchao Zhang$^1$, Yongyuan Zhu$^4$ and Heyuan Zhu$^1$}
\affiliation{$^1$Department of Optical Science and Engineering and Key Laboratory of Micro and Nano Photonic Structures (Ministry of Education), Fudan University, Shanghai 200433, China\\
$^2$Ningbo Institute of Materials Technology and Engineering, Chinese Academy of Sciences, Ningbo 315201, China\\
$^3$State Key Laboratory of ASIC and System, Institute of Advanced Nanodevices,School of Microelectronics, Fudan University, Shanghai 200433, China.\\
$^4$Nanjing University, National Laboratory of Solid State Microstructure, Nanjing 210093, China}

\begin{abstract}
In scaling of transistor dimensions with low source-to-drain currents, 1D semiconductors with certain electronic properties are highly desired. We discover three new 1D materials, SbSeI, SbSI and SbSBr with high stability and novel electronic properties based on first principles calculations. Both dynamical and thermal stability of these 1D materials are examined. The bulk-to-1D transition results in dramatic changes in band gap, effective mass and static dielectric constant due to quantum confinement, making 1D SbSeI a highly promising channel material for transistors with gate length shorter than 1 nm. Under small uniaxial strain, these materials are transformed from indirect into direct band gap semiconductors, paving the way for optoelectronic devices and mechanical sensors. Moreover, the thermoelectric performance of these materials is significantly improved over their bulk counterparts. Finally, we demonstrate the experimental feasibility of synthesizing such atomically sharp V-VI-VII compounds. These highly desirable properties render SbSeI, SbSI and SbSBr promising 1D materials for applications in future microelectronics, optoelectronics, mechanical sensors, and thermoelectrics.
\end{abstract}

\maketitle

As transistors are scaled down continually following Moore's law, a scaling limit of 5 nm arises from short channel effect that increases source-to-drain tunneling \cite{Lundstrom2003}. Recently, 2D MoS$_2$ transistors with 1 nm gate lengths have been made using 1D single-walled carbon nanotube gate \cite{Desai2016}. The introduction of semiconductors with large band gap, heavy effective mass, and low dielectric constant, like MoS$_2$, can lead to further scaling of transistor dimensions with low direct source-to-drain tunneling currents \cite{Martinez2009,Cho2015}. Considering shrinking transistor dimensions, 1D semiconductors with such properties are highly sought.

1D crystals represent the sharpest form of crystalline solids, and exhibit many exotic properties that are absent in their bulk counterpart, such as high mechanical strength and flexibility \cite{Dresselhaus2003}, promising white-light emission \cite{Yuan2017}, and enhanced piezoelectric \cite{Agrawal2011} and thermoelectric performance \cite{Hicks1993}. Carbon nanotube transistors have been fabricated for decades \cite{Tans1998,Bachtold2001,Javey2003}. However, metallic and semiconducting carbon nanotubes are typically grown together as bundles, and for well-controlled transport properties, separating tubes with suitable electronic properties remains a challenge, which limits their industrial feasibility \cite{Krupke2003}. Thus the identification of experimentally feasible 1D materials for applications in electronic devices is of utmost importance. Here, we show an \textit{ab initio} calculation evidence of three novel 1D V-VI-VII compounds, SbSeI, SbSBr and SbSI, with high stability and desired electronic properties.

Before the discovery of graphene \cite{Novoselov2004}, monolayer MoS$_2$ \cite{Mak2010}, and black phosphorene \cite{Li2014a}, studies on their bulk counterparts are well ahead of the 2D form. Similarly, properties of bulk V-VI-VII compounds are well known much earlier than their 1D form. Bulk V-VI-VII semiconductor has a orthorhombic structure with the space group of $Pna2_1$, and crystallises in 1D chains which are held together by van der Waals interaction \cite{Amoroso2016}. Bulk V-VI-VII semiconductors are earth abundant materials \cite{Demartin2015,Demartin2016}, and with chained structures, these materials may be good precursors for 1D crystals by exfoliation or mechanical cleavage \cite{Dai2015}. However, due to the lack of knowledge on the properties of 1D V-VI-VII compounds, little research has been conducted on the isolation of the chain. Historically, bulk V-VI-VII compounds attracted tremendous interest in the 1960's due to promising ferroelectric properties \cite{Fatuzzo1962,Berlincourt1964,Nitsche1964,Fridkin1967,Grekov1969,Furman1973}. The last five years have witnessed the renaissance of these materials for photovoltaic applications beyond perovskites \cite{Hahn2012,Kwolek2015,Butler2015,Butler2016,Ganose2016,Ganose2016a,Shi2016,Tablero2016}, because (i) orienting crystal growth perpendicular to the substrate sustains excellent carrier transport along the chains, (ii) benign grain boundaries parallel to the chains are free of dangling bonds and hence cause little recombination loss \cite{Zhou2015a}, and (iii) needle-like crystals aligned in the translational direction of the growth exhibit better photovoltaic response than their higher dimensional counterparts \cite{Wibowo2013}. For applications in field-effect transistors, the Achilles' heel of bulk V-VI-VII semiconductors is smaller band gap, lighter effective mass and much higher dielectric constant than 2D MoS$_2$ \cite{Butler2015,Ganose2016a,Amoroso2016,Desai2016}, which seriously reduces their potential. Interestingly, we show for the first time that the bulk-to-1D transition is accompanied by an abrupt switch in band gap, effective mass and dielectric constant, distinguishing 1D SbSeI as a promising channel material for next generation field-effect transistors.

\begin{figure}
\includegraphics[width=0.45\linewidth]{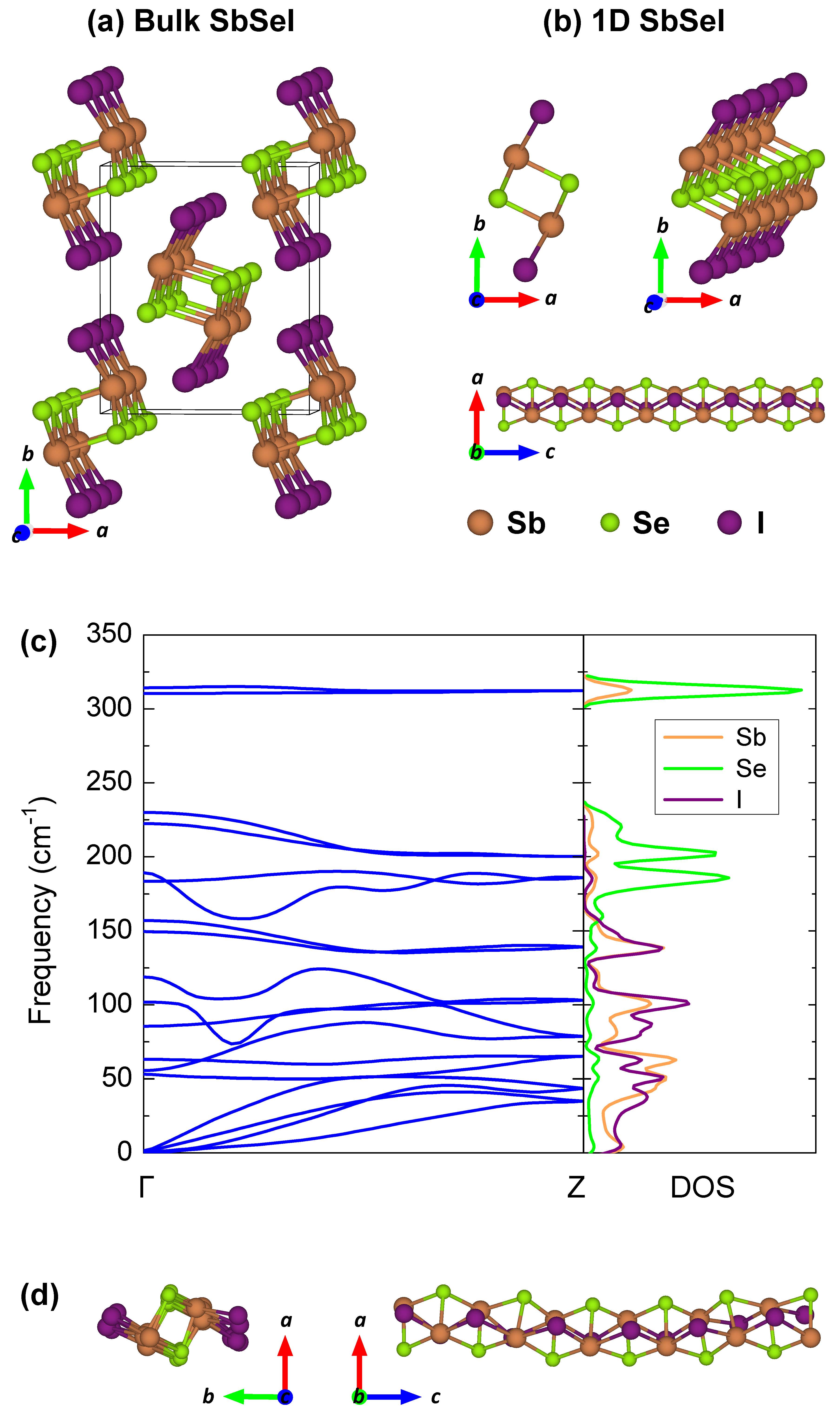}
\caption{Geometry structure of (a) bulk and (b) 1D SbSeI. (c) Phonon dispersion and phonon density of states (DOS) of 1D SbSeI. (d) 1D SbSeI supercell in molecular dynamics simulations at 300 K after 8 ps.}
\label{1} 
\end{figure}

V-VI-VII compounds have similar structures, thus we take the model of SbSeI hereafter. The PBE-D2 optimized structure of bulk SbSeI is shown in Figure~\ref{1}(a). The computed lattice constants of SbSeI, SbSI and SbSBr are listed in Table~S1 in the Supporting Information, all in good agreement with previous result \cite{Butler2016}, indicating that the theoretical methods chosen for this system is reliable. The crystal structure of isolated 1D SbSeI is shown in Figure~\ref{1}(b). The stability of atomically sharp 1D SbSeI is an important issue to be addressed. A material is dynamically stable when no imaginary phonon frequencies exist \cite{Zhang2012}. The phonon dispersion shows real frequencies across the Brillouin zone in Figure~\ref{1}(c), indicating that these structures are kinetically stable at 0 K. V-VI-VII compounds are phase transition materials, thus, to examine the thermal stability of their 1D counterparts, we carry out first-principles molecular dynamics simulations in canonical ensemble. At 300 K, a 1$\times$1$\times$6 supercell is used with a time step of 2 fs. A snapshot of atomically sharp SbSeI is shown in Figure~\ref{1}(d) with a total simulation time of 8 ps. No bond is broken and the structure integrity is well kept as its equilibrium structure at 0 K, suggesting good thermal stability.

\begin{figure}
\includegraphics[width=0.4\linewidth]{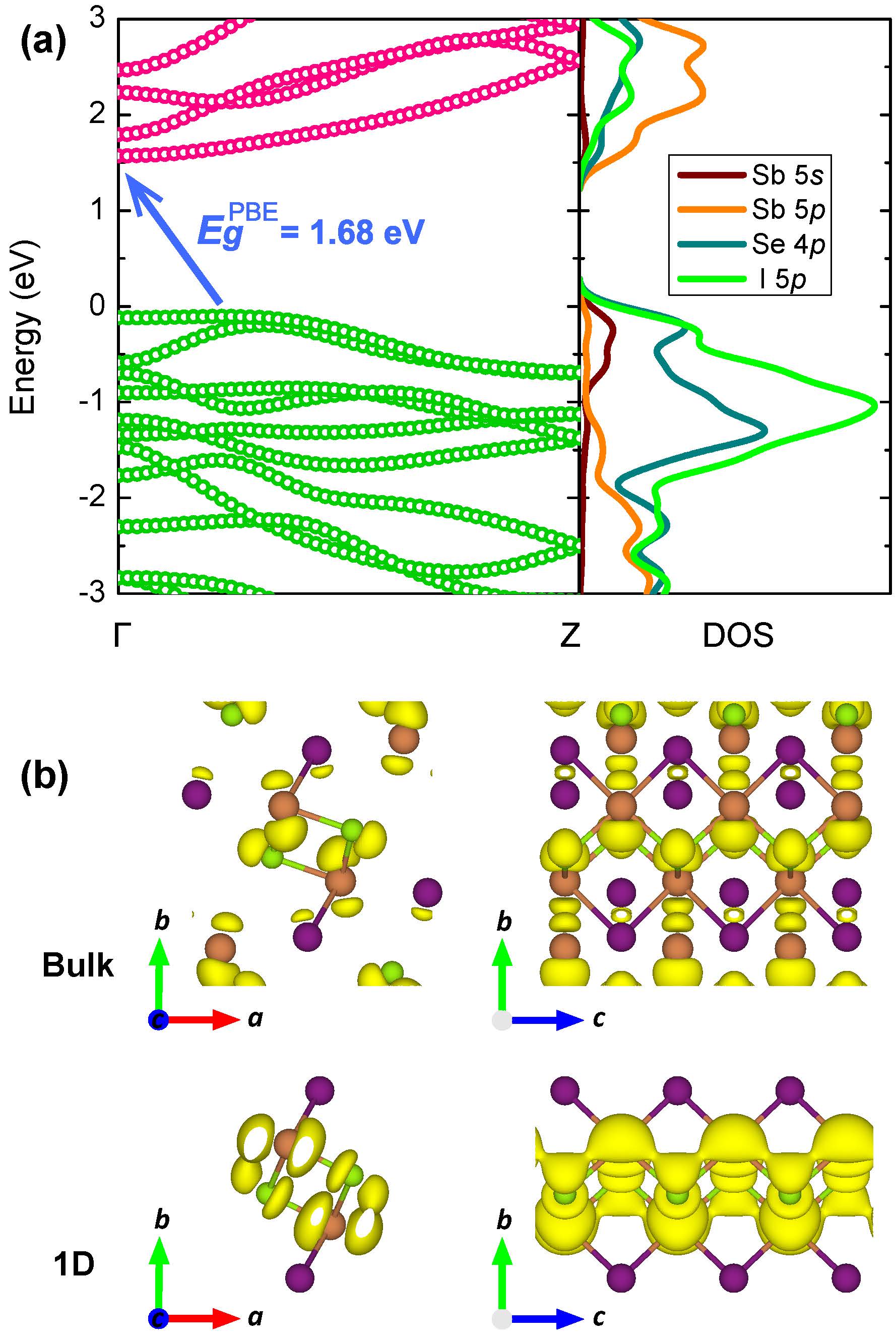}
\caption{(a) Band structure and DOS of 1D SbSeI. (b) Isosurface plots of the charge density of conduction band minimum of bulk SbSeI at the S point and 1D SbSeI at the $\Gamma$ point, with an isosurface level of 0.004.}
\label{2} 
\end{figure}

\begin{table}
\centering
\caption{Calculated PBE-D2 and HSE06 band gap, hole and electron effective mass, and static dielectric constant along the $c$ direction of 1D SbSeI, SbSI and SbSBr. Previous theoretical results of bulk SbSeI, SbSI, SbSBr and 2D MoS$_2$ are also given in parentheses for comparison.}
\begin{tabular}{cccccccc}
\hline
 & & $E_g^{\textsc{PBE}}$ (eV) & $E_g^{\textsc{HSE}}$ (eV) & $m_h$ ($m_0$) & $m_e$ ($m_0$) & $\epsilon$ ($\epsilon_0$) \\
\hline
 SbSeI & 1D & 1.68 & 2.41 & 1.67 & 1.19 & 1.86 \\
 & bulk & 1.33 & (1.86 \cite{Butler2016}) & (0.58 \cite{Butler2016}) & (0.54 \cite{Butler2016}) & 14.70 \\
 SbSI & 1D & 2.08 & 2.86 & 0.34 & 0.65 & 1.79 \\
 & bulk & 1.65 & (2.11 \cite{Butler2016}) & (0.57 \cite{Butler2016}) & (0.43 \cite{Butler2016}) & 10.56 \\
 SbSBr & 1D & 2.06 & 2.86 & 5.76 & 0.55 & 1.61 \\
 & bulk & 1.69 & (2.31 \cite{Butler2016}) & (0.64 \cite{Butler2016}) & (0.51 \cite{Butler2016}) & 13.81 \\
 MoS$_2$ & 2D & (1.60 \cite{Ramasubramaniam2012}) & (2.05 \cite{Ramasubramaniam2012}) & (0.54 \cite{Ramasubramaniam2012}) & (0.60 \cite{Ramasubramaniam2012}) & (4.62 \cite{Ramasubramaniam2012}) \\
\hline
\end{tabular}
\label{t1}
\end{table}

We then examine the PBE-D2 band structure of bulk SbSeI in Figure~S1. Bulk SbSeI is an indirect band gap semiconductor with the valence band maximum at the (0,0.429,0) point and the conduction band minimum at the S (0.5,0.5,0) point. The PBE-D2 computation gives a band gap of 1.33 eV. The computed band structure and density of states (DOS) of 1D SbSeI are shown in Figure~\ref{2}(a). A new band gap of 1.68 eV is formed, with the valence band maximum at the (0,0,0.111) point and the conduction band minimum at the $\Gamma$ point. We also calculate the band gap of 1D SbSeI using the HSE06 functional, as listed in Table~\ref{t1}. SbSeI is transformed into an 1D wide-band-gap semiconductor with the computed HSE06 band gap of 2.41 eV, which is much larger than that of bulk SbSeI \cite{Butler2016}. A similar trend is observed in 1D SbSI and SbSBr, as shown in Table~\ref{t1}. 

To understand the underlying mechanism, we compare the charge density of conduction band minimum of bulk SbSeI at the S point and 1D SbSeI at the $\Gamma$ point in Figure~\ref{2}(b). The inter-chain interaction is completely absent in 1D SbSeI, making the charge density of the conduction band minimum states quite isolated along $c$, as also for the charge density of the valence band maximum states in Figure~S2. According to quantum confinement theory, there are relatively few quantum states for the direction normal to the covalently bonded 1D axis, and the energy bands split into discrete sub-bands. As the wire diameter decreases, the valence sub-bands move down one by one and the conduction sub-bands move up one by one \cite{Dresselhaus2007}. The same effect is observed in the DOS of bulk and 1D SbSeI in Figure~S3. Therefore, similar to the bulk-to-2D transition in monolayer MoS$_2$ \cite{Kuc2011} and antimonene \cite{Zhang2015e}, the band gaps of 1D V-VI-VII compounds are much larger than their bulk counterparts. In addition, the lack of the inter-chain van der Waals interactions may also play a role of perturbation in the band gap transition.

In scaling of transistor dimensions with low source-to-drain currents, semiconductors with large band gap, heavy effective mass, and low static dielectric constant, are highly desired. Indeed, quantum confinement not only leads to a large band gap in 1D SbSeI, but also results in an abrupt change in effective mass and dielectric constant. We compare the effective masses and dielectric constants of all studied V-VI-VII compounds in Table~\ref{t1}. Similar to Si nanowires, in which the effective mass increases as the diameter decreases \cite{Gnani2007}, the main effects of quantum confinement are the shift of the subband edges and the change of the dispersion relationship. Thus the effective masses change dramatically in the bulk-to-1D transition. Large static dielectric constants $\epsilon$ are observed in bulk V-VI-VII materials, because a halide with $ns^2$ ions ($e.g.$ Sb$^{3+}$, Sn$^{2+}$) usually has enhanced cation-anion hybridization that increases lattice polarization \cite{Ganose2016,Shi2016}. However, as the size approaches several nanometers, reduction in $\epsilon$ becomes significant \cite{Tsu1997}, because electrons and holes are bound in quantized states. The static dielectric constants of 1D V-VI-VII materials along the covalently bonded 1D axis become much smaller than their bulk counterparts. It is worth to mention that, compared to monolayer MoS$_2$ \cite{Desai2016,Ramasubramaniam2012}, the band gap, effective mass and dielectric constant of bulk SbSeI are less desirable, while the larger band gap, larger effective mass and smaller dielectric constant make 1D SbSeI a more promising channel material than MoS$_2$.

\begin{table}
\centering
\caption{Calculated 1D elastic modulus, deformation potential, carrier mobility and relaxation time of 1D SbSeI, SbSI and SbSBr at 300 K.}
\begin{tabular}{cccccccc}
\hline
 & carrier type & $C^{1D}$ ($\times10^{-9}$N) & $E_d$ (eV) & $\mu_{C}$ (cm$^2$/Vs) & $\tau$ (fs) \\
\hline
 SbSeI & hole & 23.88 & -2.83 & 6.84 & 6.50 \\
 & electron & 23.88 & 1.88	& 25.69 & 17.40 \\
 SbSI & hole & 13.54 & -3.80 & 23.41 & 4.53 \\
 & electron & 13.54 & 1.53 & 54.55 & 20.26 \\
 SbSBr & hole & 12.59 & -2.65 & 0.64 & 2.10 \\
 & electron & 12.59 & 1.26 & 96.68 & 30.01 \\
\hline
\end{tabular}
\label{t2}
\end{table}

The applications of 1D semiconductors in field-effect transistors are governed by the carrier mobility as well. Although 1D SbSeI has large effective mass, the acoustic phonon limited carrier mobility is also determined by the deformation potential and the 1D elastic modulus \cite{Beleznay2003,Xi2012,Qiao2014,Zhang2016}. These data and the calculated mobility of 1D V-VI-VII compounds are summarized in Table~\ref{t2}. The electron mobilities of these 1D materials at 300 K are much higher than the hole mobilities. The good electron mobilities, 25.69 cm$^2$/Vs in SbSeI, 54.55 cm$^2$/Vs in SbSI and 96.68 cm$^2$/Vs in SbSBr respectively, indicate their high performance in microelectronic devices.

For optoelectronic and photovoltaic devices, indirect band gap semiconductors often suffer from poor efficiency. However, the difference between direct and indirect band gaps of 1D SbSeI is around 0.02 eV. Therefore, similar to bulk SbSeI \cite{Butler2016}, 1D SbSeI may combine the benefits of direct and indirect band gap materials: The optical absorption remains strong because of the small energy separation between different band edges; while electron-hole recombination is suppressed, since carriers can be thermalized to basins of different electron momentum. Furthermore, by applying a small tensile strain of 3\%, 1D SbSeI undergoes an indirect-to-direct band gap transition (for details, see Figure~S4 in the Supporting Information). Moreover, 1D materials can tolerate relatively large deformations \cite{Agrawal2011}, and this direct band gap persists under larger tensile strain, which implies that these 1D materials are promising in optoelectronic devices and mechanical sensors.

\begin{figure}
\includegraphics[width=0.4\linewidth]{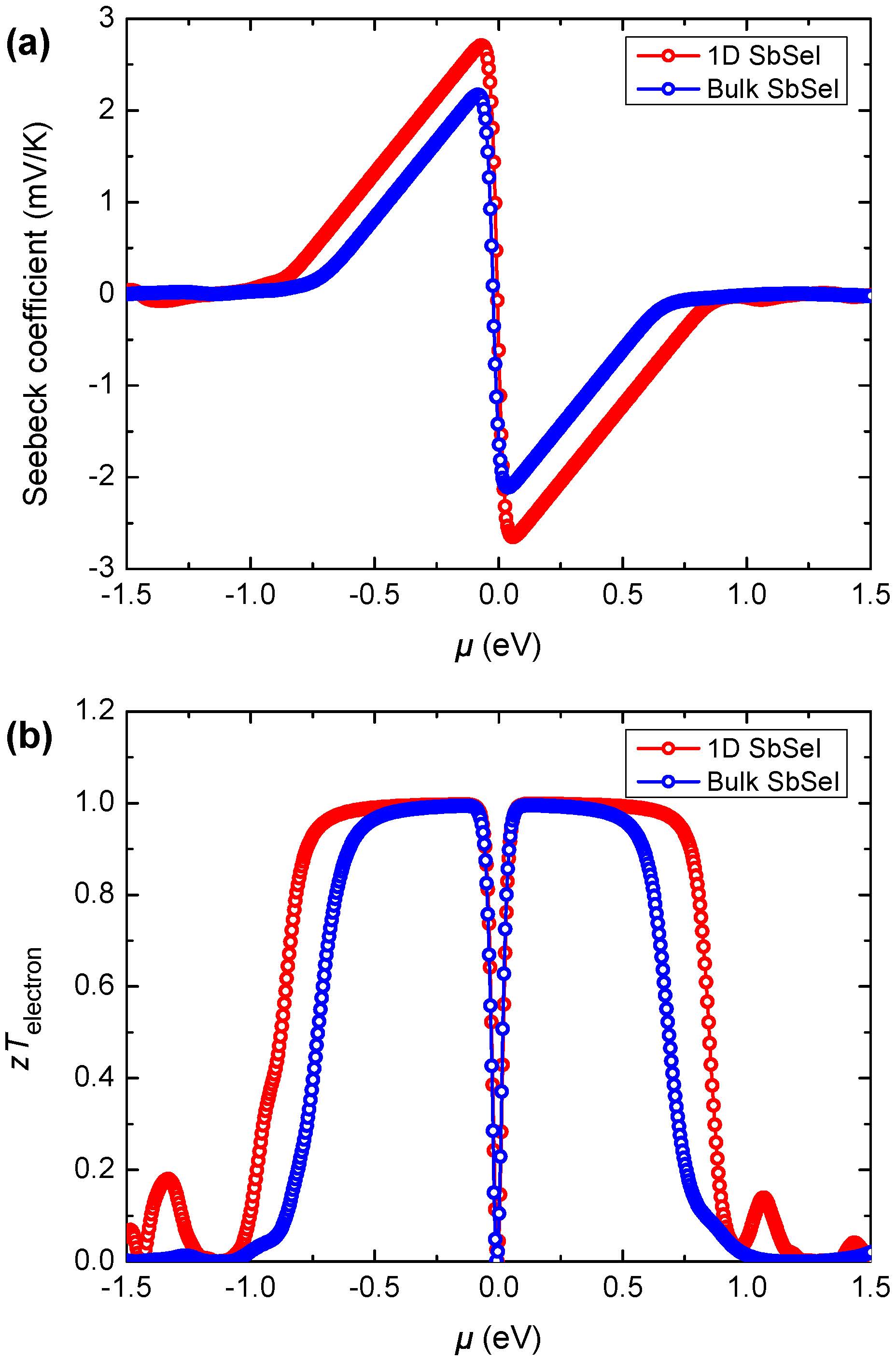}
\caption{(a) Seebeck coefficient and (b) electronic figure of merit $zT$ of 1D SbSeI at 300 K.}
\label{3} 
\end{figure}

Furthermore, with decreasing dimensionality to 1D, the thermoelectric efficiency can be improved dramatically, since the electrons are confined in a single dimension, and the phonons are easily scattered by the boundary \cite{Hicks1993}. Figure~\ref{3}(a) shows the calculated Seebeck coefficient of 1D and bulk SbSeI, where a negative $\mu$ means p-type doping, while a positive $\mu$ means n-type doping. The Seebeck coefficient $S$ of 1D SbSeI is much larger than that of bulk SbSeI, indicating higher thermoelectric performance. Typically, $S$ is large at the edge of heavy band \cite{MehdizadehDehkordi2015}, and 1D SbSeI possesses larger effective mass due to quantum confinement, thus the $S$ increases significantly. Even bulk SbSeI has an ultralow lattice thermal conductivity along the $c$ direction at 300 K, as shown in Table~S2 in the Supporting Information. Thereafter we only calculate the electronic figure of merit, which quantifies the electron-limited thermoelectric efficiency. Figure~\ref{3}(b) presents the calculated thermoelectric figure of merit $zT$ for 1D and bulk SbSeI at 300 K using semiclassical Boltzmann transport theory \cite{Madsen2006}. High thermoelectric performance of 1D SbSeI is observed in a wider doping range than bulk SbSeI. Considering the enhanced electrical transport and suppressed phonon transport in 1D nanostructures, the thermoelectric performance of 1D SbSeI can reach its maximum electron limited $zT$ more easily than its bulk counterpart.

Although these materials exhibit novel properties for various applications, their experimental feasibility has yet to be confirmed. To examine the feasibility of isolation of the 1D chain, we calculate the cleavage energy of SbSeI. The ideal cleavage energy of 0.52$\times10^{-9}$ J/m is much lower than that of graphite \cite{Wang2015c}, indicating that the isolation is experimentally attainable. Besides the exfoliation of bulk V-VI-VII crystals, the synthesis of 1D V-VI-VII materials by hydrothermal method or sonochemical method may be experimentally feasible as well, note that their nanowires with diameters of 10-50 nm and lengths up to several micrometers have been synthesized using these methods \cite{Wang2001,Yang2001,Nowak2008,Nowak2009}. Finally, we predict the Raman frequencies for bulk and 1D V-VI-VII compounds in Table~S3 in the Supporting Information as a reference for characterization.

In summary, we predict for the first time three novel 1D materials, SbSeI, SbSI and SbSBr. Both the dynamical stability and the thermal stability of these atomically sharp materials are confirmed. During the bulk-to-1D transition, an abrupt change in band gap, effective mass and static dielectric constant due to quantum confinement makes 1D SbSeI highly desirable in next-generation field-effect transistors. A good electron mobility is also observed in these 1D V-VI-VII compounds. Furthermore, 1D SbSeI experiences an indirect-to-direct band gap transition under a small tensile strain of 3\%, indicating promising applications in optoelectronic devices and mechanical sensors. A significant increase in thermoelectric performance can be expected in 1D nanostructures as well. Lastly, the syntheses of these materials by exfoliation, hydrothermal method and sonochemical method can be experimentally feasible. We believe that the fabrication of these novel materials in the laboratory will be accomplished in the near future.

\subsection*{Experimental section}

See in the Supplementary Information.

\section*{Acknowledgement}

This work is supported by the National Natural Science Foundation of China under Grants No. 11374063 and 11404348, and the National Basic Research Program of China (973 Program) under Grant No. 2013CBA01505.


%

\end{document}